\documentclass[11pt]{article}
\usepackage{times}
\usepackage{latexsym}
\usepackage[T1]{fontenc}
\usepackage[utf8]{inputenc}
\usepackage{inconsolata}%
\usepackage[table,dvipsnames]{xcolor}
\usepackage{acl}
\usepackage{multirow}
\usepackage[ruled, lined, linesnumbered, commentsnumbered, longend]{algorithm2e}
\usepackage{algpseudocode}
\usepackage{comment}
\usepackage{caption}
\usepackage{array}
\usepackage{float}
\usepackage{tabularx}
\usepackage{makecell}
\usepackage{mathtools}
\usepackage{amsmath}
\usepackage{graphicx} 
\usepackage{tabularx}
\usepackage{makecell}
\usepackage{todonotes}
\usepackage{multirow}
\usepackage{caption}
\usepackage{balance}
\usepackage{lipsum}
\usepackage{array}
\usepackage{adjustbox}
\usepackage{comment}
\usepackage{subfigure}
\usepackage{color, colortbl}

\DeclareMathOperator*{\argminA}{arg\,min}
\begin{document}
\title{A Counterfactual Explanation Framework for Retrieval Models}

\author{ Bhavik Chandna$^\dagger$  ~   Procheta Sen$^\clubsuit$ \\
     University of San Diego, USA$^\dagger$ \\
     University of Liverpool, United Kingdom.$^\clubsuit$\\
        {\tt bhavikchandna@gmail.com$^\dagger$ procheta.sen@liverpool.ac.uk$^\clubsuit$} 
 }




\maketitle
\begin{abstract}
 Explainability has become a crucial concern in today's world, aiming to enhance transparency in machine learning and deep learning models. Information retrieval is no exception to this trend. In existing literature on explainability of information retrieval, the emphasis has predominantly been on illustrating the concept of relevance concerning a retrieval model. The questions addressed include why a document is relevant to a query, why one document exhibits higher relevance than another, or why a specific set of documents is deemed relevant for a query. However, limited attention has been given to understanding why a particular document is not favored (e.g., not within top-K)  with respect to a query and a retrieval model. In an effort to address this gap, our work focuses on the question of what terms need to be added within a document to improve its ranking. This, in turn, answers the question of which words in the document played a role in not being favored by a retrieval model for a particular query. We use a counterfactual framework to solve the above-mentioned research problem.
To the best of our knowledge, we mark the first attempt to tackle this specific counterfactual problem (i.e. examining the absence of which words can affect the ranking of a document). Our experiments show the effectiveness of our proposed approach in predicting counterfactuals for both statistical (e.g. BM25) and deep-learning-based models (e.g. DRMM, DSSM, ColBERT, MonoT5). The code implementation of our proposed approach is available in \url{https://anonymous.4open.science/r/CfIR-v2}.
\end{abstract}

\section{Introduction}
The requirement of transparency of Artificial Intelligence (AI) models has made explainability crucial, and this applies to Information Retrieval (IR) models as well \cite{Anand}. The target audience plays a significant role in achieving explainability for an IR model, as the units of explanation or questions may differ based on the end user. For instance, a healthcare specialist, who is a domain expert but not necessarily an IR specialist, might want to understand the reasons behind a ranked suggestion produced by a retrieval model in terms of words used \cite{exs}. On the other hand, an IR practitioner may be more interested in understanding whether different IR axioms are followed by a retrieval model or not \cite{bondarenko:2022:sigir:AxiomRetrievalIRAxioms}.

This study focuses on the perspective of IR practitioners. To be more specific, we introduce a counterfactual framework designed for retrieval models, catering to the needs of IR practitioners. Existing literature in explainable IR (ExIR) addressed questions like why a particular document is relevant with respect to a query \cite{exs}, between a pair of documents why one document is more relevant to the query \cite{penha:2022:sigir:generatePointPair} compared to the other and why a list of documents relevant to a query \cite{IXS}. Broadly speaking, all the above-mentioned questions mainly focus on explaining the relevance of a document or a list of documents from different perspectives. 

However, there is limited attention to explain the question like the absence of which words renders a document unfavorable to a retrieval model (i.e. not within top-K) remains unexplored. The above-mentioned explanation can give an idea to an IR practitioner about how to modify a retrieval model. For example, if it is observed that a retrieval model (e.g. especially neural IR models \cite{10.1145/3397271.3401280})  does not favor documents because of not having certain gender specific words then the setting of the retrieval model needs to be debiased. 

In many realistic retrieval settings—such as patent search, legal case retrieval, and clinical information access—users and IR engineers frequently need to understand not only why a document was retrieved, but also why a potentially relevant document failed to appear in the top-K results. Missing a relevant document can have legal, financial, or safety-critical implications. In such environments, stakeholders require per-document, contrastive explanations that specify what information was absent from the document and prevented it from being retrieved. 

With the motivation described above, the fundamental research question which we address in this research work is  \textbf{RQ1:} `\textit{What are the terms that should be added to a document which can push the document to a higher rank with respect to a particular retrieval model?}'

We would like to note that we have framed \textbf{RQ1} as a counterfactual setup in our research scope. Similar to existing research in counterfactual explanations in AI \cite{kanamori2021,van2021}, we also attempt to change the output of model with the provided explanations (i.e. change the rank of a document in IR models). 
Our experimental results show that on an average in $70\%$ cases the solution provided by the counterfactual setup improves the ranking of a document with respect to a query and a ranking model. 

\paragraph{Our Contributions }The main contributions of this paper are as follows.

\begin{itemize}
    \item Propose a model-agnostic novel counterfactual framework for retrieval models.

    \item Estimated a set of terms that can explain why a document is not within top-K with respect to a query and a retrieval model. 

    \item Provide a comprehensive analysis with existing state-of-the-art IR models.
\end{itemize}

The rest of the paper is organized as follows. Section \ref{sec:relwork} describes Related work. Section \ref{sec:cf} describes the counterfactual framework used in our work, Section \ref{sec:setup} describes the experimental setup and Section \ref{sec:results} discuss about results and ablation study. Section \ref{sec:conc} concludes with this paper.
\section{Related Work}\label{sec:relwork}

\paragraph{Counterfactual Explanations} The xAI field gained significant momentum with the development of the Local Interpretable Model-agnostic Explanations (LIME) method \cite{lime}, which offers a way to explain any classification model. While models like LIME explain why a model predicts a particular output, counterfactual explainers address the question of what changes in input features would be needed to alter the output. Counterfactual xAI was first brought into the limelight in early 2010s with seminal work of \citet{pearl2018theoretical}. The study in \citet{karimi2020model} provided a practical framework named Model-Agnostic Counterfactual Explanations (MACE) for any model. Later series of models \cite{kanamori2021,van2021,parmentier2021,carreira2021,pawelczyk2022exploring,icml} were proposed for counterfactual explanation based on different optimization frameworks. In our research scope, we use Counterfactual Explanation framework proposed in \cite{mothilal2020dice} (explained in detail in Section \ref{sec:cf}).

\paragraph{Explainability in IR}\label{sec:expir}

\textbf{Pointwise Explanations} shows the important features responsible for the relevance score predicted by a retrieval model for a query-document pair. Popular techniques include locally approximating the relevance scores predicted  by the retrieval model using a regression model \citep{exs}.

\textbf{Pairwise Explanations} predict why a particular document was favored by a ranking model compared to others. 
The work in \cite{xu2024counterfactual} proposed a counterfactual explanation method to compare the ranking of a pair of documents with respect to a particular query. 

\textbf{Listwise Explanations} focus on explaining the key features for a ranked list of documents and a query. Listwise explanations \cite{yu2022towards, IXS} aim to capture a more global perspective compared to pointwise and pairwise explanations. The study in \cite{IXS} proposed an approach which combines the output of different explainers to capture the different aspects of relevance. The study in \cite{yu2022towards} trained a transformer model to generate explanation terms for a query and a ranked list of documents.

\textbf{Generative Explanations} \cite{singh:2020:fat:intentmodel,IXS} generally leverage natural language processing to create new text content, like summaries or justifications, that directly address the user's query and information needs. Model-agnostic approaches \cite{singh:2020:fat:intentmodel} have been proposed to interpret the intent of the query as understood by a black box ranker.

From the above mentioned category of explanations in IR, we focus on pointwise explanation in our research scope. In pointwise explanation, rather than explaining what are the words which are relevant in a document for a particular query we address the research question what are the words which are required to improve the ranking of the document with respect to a query. 
\begin{figure*}
    \centering
\includegraphics[width=0.7\textwidth]{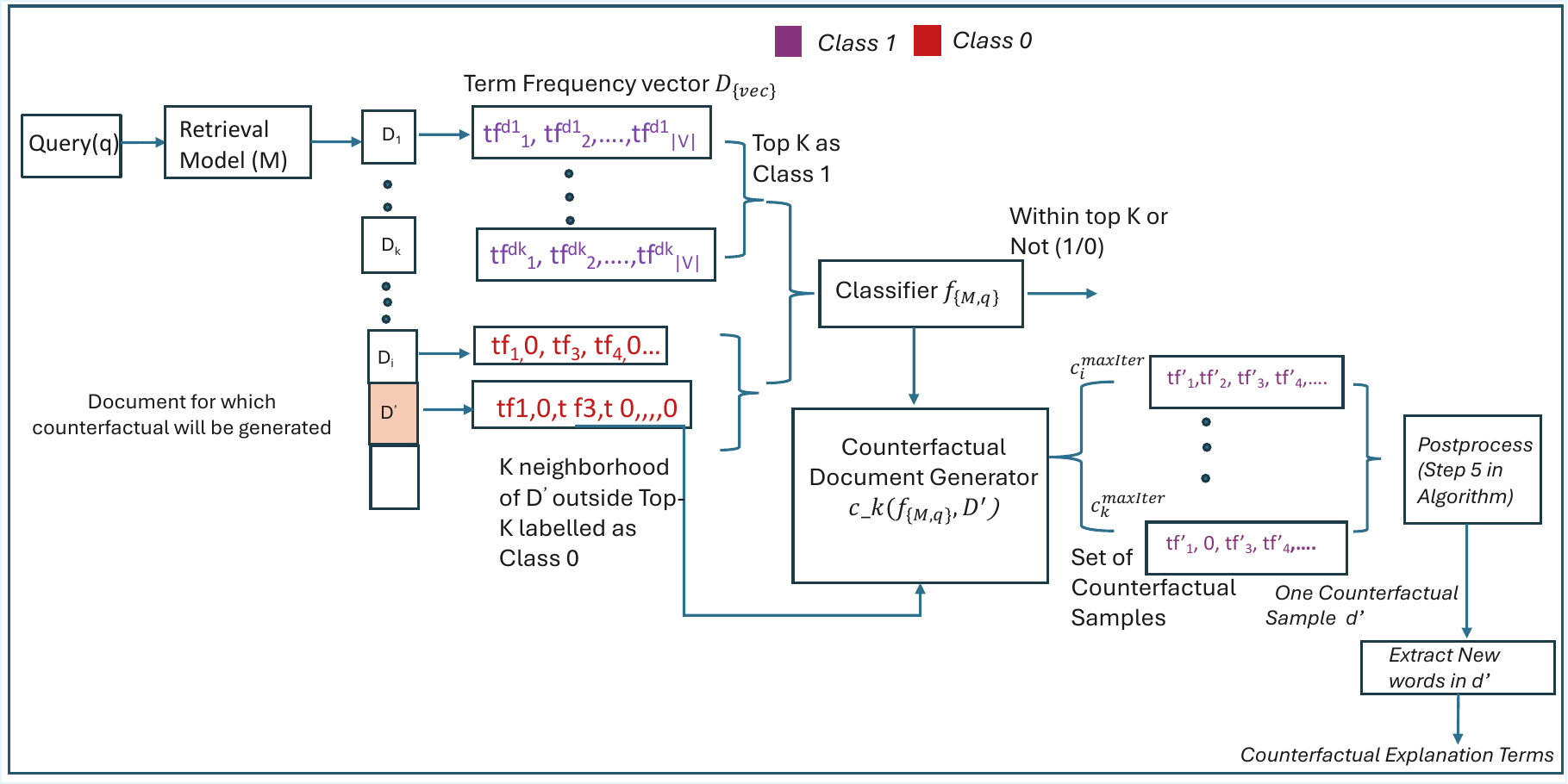}
    \caption{Schematic Diagram for Counterfactual Explanation Framework (CFIR)}
    \label{fig:countersetup}
\end{figure*}
\paragraph{Search Engine Optimization} techniques \cite{seo1, seo2} generally uses different features like commercial cost, links to optimize the performance of the search engine. A major difference of the work in \cite{seo1,seo2} with our work is we only consider the words present in a document as a feature. Our objective is to improve the ranking of a particular document concerning a specific query and a retrieval model rather than improving the ranking of a document concerning any query belonging to a particular topic.  

\section{Counterfactual Framework for Information Retrieval (CFIR)}\label{sec:cf}
\paragraph{Problem Statement} 
Let $d$ represents a target document that does not appear in the top-$K$ retrieved results of a query $q$ and retrieval model $M$. The objective in CFIR is to identify a set of terms ${w_i}$ which, when added to $d$, improve its ranking with respect to $q$ and model $M$.

 The above mentioned setup for CFIR is formally defined in Equation~\ref{eq:cfir} where \textit{CFIR}, employs a counterfactual document generator $c_{k}(f_{\{M,q\}}, d)$ which takes as input a classifier $f_{M,q}$  and the document $d$  to construct an counterfactual document $d'$ such that $d'$ is likely to get a higher rank (within top-$K$) than $d$ for model $M$ and query $q$. The objective of $f_{\{M,q\}}:R^{|V|} \rightarrow \{0,1\}$ ( where $V$ is the vocabulary, described in detail in Section \ref{sec:classifier}) is to predict given a query $q$ and a retrieval model $M$ if a particular document $d$ will be within top-K or not. 
The counterfactual explanation is defined as the set of words present in $d'$ but not in $d$ (i.e. output of Equation \ref{eq:cfir}). 

\scriptsize
{\begin{align}
 \nonumber CFIR(q,M,d)&=c_{k}(f_{\{M,q\}},d)-d \\
    &=d'-d =\cup_{i=1}^m\{w_i\}
    \label{eq:cfir}
\end{align}}
\normalsize

\subsection{Building Classifier ($\mathbf{f_{\{M,q\}}}$)}\label{sec:classifier}  

Similar to existing xAI \cite{lime} approaches, the classifier $f_{\{M,q\}}$ in our research scope essentially locally approximates the behavior of a retrieval model $M$, for a query $q$ and a subset of documents retrieved for the query $q$. 
However, in contrast to the regression model in \cite{lime}, we build a binary classification model to predict whether a document $d$ will be ranked within the top-$K$ results or not for a specific query $q$ and retrieval model $M$.

For each query q, we build a classifier which predicts whether a document will be retrived in top-k or not with respect to a Model M. To build this classifier we take top K documents from the

In the classifier setup, the top-$K$ documents for a query $q$ and retrieval model $M$ represent class $1$ and any other document not belonging to this class represents class $0$. Theoretically speaking, if a corpus had $N$ number of documents, then there will be $N-K$ documents which should have class label $0$ and $N-K$ is a very large number in general which can cause class imbalance issue.  

To avoid this issue, we choose only K documents from the set $N-K$. Out of this $K$  documents  we use the $x$  number of documents for which we want to generate the explanations and then we choose randomly selected $K-x$ documents from the N-K set. $K$ serves as a predefined threshold, typically set to values such as $10$, $20$, or $30$. For $f_{\{M,q\}}$, each document $d$ is represented as a word term frequency based feature vector, denoted as $d_{vec}$.

Formally, \textbf{Feature Vector for Classifier $\mathbf{f_{\{M,q\}}}$} \label{sec:fv} is represented as $d_{vec}=\{tf_1^d, tf_2^d\ldots,tf_{|V|}^d\}$ where $tf_i^d$ represents the term frequency of the word $w_i$ in $d$. Using all the words from all the documents retrieved for a query to construct the vocabulary set $V$ can pose challenges. Consequently, we take the union of the most significant $n$ words from each document $d$ using a function named $Imp(d)$ (explained in detail in Section \ref{sec:setup}) to construct $V$. 
   $ V=\cup_{i=1}^{K}\{\cup_{j=1,w_j \in Imp(d_i)}^{n}{w_j}\}$.
Appendix \ref{ap:classifier} depicts a step-by-step algorithm to construct the feature vector for the classifier and Figure \ref{fig:exampleclass} in Appendix \ref{ap:classifier} shows one sample feature vector for the classifier.
\paragraph{Counterfactual Document Generator $\mathbf{c_k(f_{M,q},d)}$} in Equation \ref{eq:cfir} 
follows an approach similar to that of \citet{mothilal2020dice}. Specifically, $c_{k}(f_{M,q}, d)$ generates $k$ candidate counterfactuals ${c_1^{maxIter}, c_2^{maxIter}, \ldots, c_k^{maxIter}}$ (where $maxIter$ is the maximum number of iterations upto which loss function is optimized) for each document $d$, from which we randomly select a single counterfactual ($d'$ in Equation \ref{eq:cfir}) that involves only insertion of new words without modifying or deleting existing ones in $d$ (step 5 in Algorithm \ref{algorithm}). We fix $k$ to a sufficiently large constant in our experiments.
%
%
%
  Similar to \cite{mothilal2020dice}, the objective of $c_{k}(f_{M,q}, d)$ is to minimize three different criteria described as follows.
\begin{itemize}
    \item \textbf{Criteria 1:} Minimizing the distance between the desired outcome $y'$ (within top-$K$) and the prediction of the classifier model $f_{\{M,q\}}$ for a counterfactual example ($c_i$).

    \item \textbf{Criteria 2:} Minimizing the distance between any generated counterfactual ($c_i$) and the original document $d$. Broadly speaking, a counterfactual example closer to the original input should be more useful for a user.
    \item \textbf{Criteria 3:} Increasing diversity between generated counterfactuals.
\end{itemize}
Based on the above-mentioned criteria the loss function to generate $c_1^{maxIter}, \ldots c_k^{maxIter}$ is described as follows.

\scriptsize
{\begin{equation}
\begin{split}
  &  \argminA_{c_1,\ldots c_k} \left( \frac{1}{k} \sum_{i=1}^k \text{yloss}(f_{M,q}(c_i),y')+ \right. \\
     &\left. \frac{\lambda_1}{k}\sum_{i=1}^k \text{dist}(c_i,d)-\lambda_2 \text{div}(c_1,\ldots , c_k) \right)
\end{split}
\label{eq:cfsetup}
\end{equation}}
\normalsize

In Equation \ref{eq:cfsetup},  $\textit{yloss}(.)$ takes care of \textbf{Criteria 1}, $\textit{dist}(c_i,d)$ takes care of the \textbf{Criteria 2} and $\textit{div}$ takes care of the \textbf{Criteria 3} as discussed above. $\lambda_1$ and
$\lambda_2$ in Equation \ref{eq:cfsetup} are hyperparameters that balance the contribution of second and third parts of loss function (i.e. controlling diversity and similarity). The detailed description of the computation of $yloss$, $dist$ and $div$ function in Equation \ref{eq:cfsetup} is given in Equations \ref{eq:loss}, \ref{eq:madp} and \ref{eq:diversity} respectively in Appendix \ref{ap:cfsetup_appendix}. The loss function in Equation \ref{eq:cfsetup} is optimized using the gradient descent method. 

%

%
Algorithm \ref{algorithm} shows step by step execution of the counterfactual document generator $c_k(f_{\{M,q\}},d)$. In Algorithm \ref{algorithm} we show how the counterfactual examples ($c_1,\ldots c_k$) are randomly initialized. The generated counterfactual examples (i.e. $c_i^{maxIter}$s) should change the prediction of classifier $f_{\{M,q\}}$ from $0$  to $1$ (i.e. modified document should be within top $K$). 
The set of words corresponding to the counterfactual explanation of $d$ are the new words that have been added to $d'_{vec}$ (i.e. feature vector representation of $d'$ in Equation \ref{eq:cfir}) compared to $d_{vec}$. Figure \ref{fig:countersetup} shows the schematic diagram for counterfactual setup with the workflow between the different components (i.e. classifier and counterfactual document generator) within it. 
\scriptsize
\begin{algorithm}
\scriptsize
\SetKwFunction{isOddNumber}{isOddNumber}
    \SetKwInOut{KwIn}{Input}
    \SetKwInOut{KwOut}{Output}
     \SetKwInOut{KwInit}{Initialization}
    \KwIn{ Classifier function: $f_{\{M,q\}}$, Feature Vector: $d_{vec} =\{tf_1,tf_2,\ldots, tf_{|V|}\}$, Number of Counterfactuals:k }
    \KwOut{ \{$d_{vec}' \in R^{|V|}$\}}
    \KwInit{
   \\
    \For {$i \leftarrow 1$ \KwTo $k$}
    {\For {$j \leftarrow 1$ \KwTo $|V|$}
   {$c_{i,j}^0=
r \sim Random(.)$ \\
\tcc {$c_{i,j}^0$ is the $j^{th}$ coordinate of $c_{i}$ at $0^{th}$ iteration}
    }
    
    }
    }
    \For {$t \leftarrow 0$ \KwTo $maxIter$}
    {\text{Compute the loss} $\frac{1}{k} \sum_{i=1}^k \text{yloss}(f_{M,q}(c_{i}^t),y)+
     \frac{\lambda_1}{k}\sum_{i=1}^k \textit{dist}(c_{i}^{t},d)-\lambda_2 \textit{div}(c_{1}^{t},\ldots , c_{k}^{t}))$ \\
     {Update $c_{i}^{t}$'s using gradient descent}
     }
    \KwRet{$d_{vec}'$},
     $d_{vec}'$ is a $|V|$ dimensional vector randomly chosen from the subset of $c_{i}^{maxIter}$'s for which $c_{i,j}^{maxIter} \geq tf_j^d ~\forall j=1,\ldots, |V|$ 
    \caption{\scriptsize CF Document Generator $c_k(f_{\{M,q\}},d)$}
    \label{algorithm}
\end{algorithm}
\normalsize
\section{Experiment Setup} \label{sec:setup}
\paragraph{Dataset}
We use three ranking datasets for our experiments: MS MARCO passage dataset for passage ranking \cite{Bajaj2016Msmarco} and MS MARCO document ranking dataset for longer documents \cite{trecdeep} and TREC Robust \cite{96071} dataset. The MS MARCO passage and document ranking datasets contain queries from Bing\footnote{\url{https://bing.com}} and the queries of TREC Robust are manually chosen. 
For each dataset, we randomly selected $100$ queries from the test set and chose $5$ documents not ranked in the top $10$ results for each query, resulting in a test set of $500$ query-document pairs. The details of the dataset are given in Table \ref{tab:dataset} in Appendix \ref{ap:stat}.

We use five different retrieval models BM25, DRMM \citet{guo}, DSSM \cite{huang}, ColBERT \citet{colbert}, MonoT5 \cite{nogueira-etal-2020-document} and Splade \cite{splade}  in  our experiments. The details of each retrieval model is given in Appendix \ref{ap:rm}.


\paragraph{\textbf{Baselines}}
To the best of our knowledge, this is the first work which attempts to provide counterfactual explanations in IR. Consequently, there exists no baseline for our proposed approach. However we have used a query word  and top-K word based intuitive baseline to compare with our proposed approach. In query word baseline ($QW$), we use query words not originally present in a document to enhance its ranking. 
For Top-K' ($Top-K'$) baseline we use the top $k'$ words extracted from top $5$ documents corresponding to a query as relevance set. Words appearing in the relevance set but not appearing in a document are added to the document to improve its ranking. For different retrieval models we have corresponding versions of $QW$ and $Top-K'$ baselines.
\begin{table*}[htb]
   \footnotesize
\resizebox{\linewidth}{!}
{
\begin{tabular}{|lc|c|c|c|c|c|c|c|c|c|}
    \hline
     \rowcolor{yellow!25}
   \multicolumn{2}{|c|}{\textbf{Model Description}}& \multicolumn{3}{c|}{\textbf{MS MARCO Passage}} &\multicolumn{3}{|c|}{\textbf{MS MARCO Document }} &\multicolumn{3}{|c|}{\textbf{Trec Robust }} \\
     \hline
     \rowcolor{gray!30}
        \textbf{Retrieval Model} & \textbf{Classifier} &  \textbf{FD(\%)} & \makecell[c]{\textbf{Avg. New} \\ \textbf{Words}} & \makecell[cc]{\textbf{Avg. Query} \\ \textbf{Overlap}}&    \textbf{FD(\%)} & \makecell[cc]{\textbf{Avg. New} \\ \textbf{Words}} & \makecell[cc]{\textbf{Avg. Query} \\ \textbf{Overlap}}&\textbf{FD(\%)} & \makecell[cc]{\textbf{Avg. New} \\ \textbf{Words}} & \makecell[cc]{\textbf{Avg. Query} \\ \textbf{Overlap}}\\
    \hline
$QW_{BM25}$&NA& 50\% & 5.61
& 100\%  & 48\% & 6.14 
& 100\% &56\%&6.12&  100\%\\
$Top-K'_{BM25}$&NA & 42\% & 11.28 
& 100\%& 40\% & 9.61
& 100\% & 41\%&12.34&100\%\\
 $CFIR_{BM25}$ &RF& 65\% & 10.64
 & 66\%   & 52\% & \textbf{16.81} 
& 56\% &\textbf{64}\%& 11.12& 57\%\\
 $CFIR_{BM25}$ &LR & \textbf{69\%} & \textbf{17.14}  
 & 58\%  & \textbf{57\%} & 14.15
& 56\% &58\%&\textbf{13.25}& 56\% \\
 \hline
 \hline
$QW_{DRMM}$ &NA& 48\%& 5.12
& 100\% & 47\%& 6.14 
& 100\%  &49\%&7.12& 100\%\\


$Top-K'_{DRMM}$ &NA& 42\% & \textbf{15.11} 
&100\%  &31\% & 14.12 
& 100\% &33\%&\textbf{16.12}&100\% \\


$CFIR_{DRMM}$ &RF & \textbf{72\%} & 11.31
& 48\%  & 56\% & 8.12
& 46\% &62\%&12.56&47\%\\


$CFIR_{DRMM}$ &LR & 68\% & 12.37
& 62\%   & \textbf{\underline{62\%}}& \textbf{14.53}
& 45\% &\textbf{65}\%&13.47& 43\% \\
\hline
\hline
$QW_{DSMM}$ &NA & 49\%&5.32
&100\%  & 45\% & 6.64  
&100\% &52\%&7.12&100\%\\


$Top-K'_{DSSM}$ &NA & $35\%$& 12.51 
& 100\%  & 32\%&  12.62
&100\% &34\%&13.14&100\%\\
$CFIR_{DSSM}$ &RF & 57\% & 11.52
& 58\%  & 46\% & 18.14 
& 57\% &\textbf{59}\%&12.46&100\% \\
$CFIR_{DSSM}$ &LR&$\textbf{62\%}$ & \textbf{15.78} 
& 54\%  & \textbf{53\%} & \underline{\textbf{18.52}}
& 63\% &58\%&\textbf{17.24}& 64\%\\ 
\hline
\hline 
$QW_{ColBERT}$ &NA & 56\% & 4.78
& 100\% & 34\% & 5.64
& 100\% & 38\% & 6.14
& 100\%\\
$Top-K'_{ColBERT}$ &NA & 48\% & \textbf{15.63}
& 100\%  & 36\% & \textbf{13.42}
& 100\% & 38\% & 11.32
& 100\%\\
$CFIR_{ColBERT}$ &RF &72\% & 12.41
& 56\% &$\textbf{72\%}$&11.05
& 49\% &$71\%$&10.35
& 52\% \\

$CFIR_{ColBERT}$ &LR   &\textbf{75}\%
& 14.12 & 61\% & 71\%&
10.23&62\%& \textbf{74}\% & \textbf{16.45}
& 65\%\\
\hline
\hline
$QW_{MonoT5}$ &NA &52\%
& 10.15 & 100\% & 54\%&
12.23& 100\% &63\%& 10.15
& 100\%\\
$Top-K'_{MonoT5}$ &NA &75\%
& \textbf{14.11} & 100\% & 68\%&
10.13&100\%& 75\% & 11.12
& 100\%\\
$CFIR_{MonoT5}$ &RF &80\%
& 12.13 & 64\% & 72\%&
11.23&61\%& 73\% & 10.95
& 66\%\\

$CFIR_{MonoT5}$ &LR  &\underline{\textbf{82}}\%
& 13.15 & 65\% & \underline{\textbf{74}}\%&
\textbf{12.23}&63\%& \underline{\textbf{75}}\% & \textbf{11.45}
& 68\%\\
\hline
\hline
$QW_{Splade}$ &NA &49\%
& 10.15 & 100\% & 51\%&
11.51& 100\% &62\%& 11.11
& 100\%\\
$Top-K'_{Splade}$ &NA &71\%
& \textbf{13.05} & 100\% & 65\%&
9.23&100\%& 74\% & 12.22
& 100\%\\
$CFIR_{Splade}$ &RF &78\%
& 11.23 & 62\% & 69\%&
12.11&60\%& 71\% & 9.81
& 65\%\\

$CFIR_{Splade}$ &LR  &\textbf{80}\%
& 12.15 & 63\% & \underline{\textbf{71}}\%&
\textbf{14.11}&64\%& \underline{\textbf{73}}\% & \textbf{10.55}
& 67\%\\
\hline
\end{tabular}}
\caption{CFIR model Performance for BM25, DRMM, DSSM, ColBERT, MonoT5 and Splade in MSMARCO Passage and Document  Collection and TREC Robust. The Best Performing Counterfactual Explanation Method for every retrieval model is boldfaced; the overall best performance across all rows is underlined. All the results reported in Table \ref{tab:Results} are statistically significant with $p < 0.05$.}
\label{tab:Results}
\end{table*}

\paragraph{Evaluation Metrics}
There exists no standard evaluation framework for exIR approaches. The three different evaluation metrics in our experiment setup are described as follows.

\vspace{3mm}
 \textbf{Fidelity (FD):} Existing xAI approaches in IR use Fidelity \cite{Anand} as one of the metrics to evaluate the effectiveness of the proposed explainability approach. Intuitively speaking, Fidelity measures the correctness of the features obtained from a xAI approach. In the context of the CFIR setup described in this work, we define this fidelity score as the number of times the words predicted by the counterfactual algorithm could actually improve the rank of a document. Let $n$ be total number of query document pairs in our test case and x be number of query document pairs for which the the rank of the document improved after adding the counterfactuals obtained from the optimization setup described in Equation \ref{eq:cfsetup}. Then the Fidelity score is mathematically defined with respect to a test dataset $D$ and retrieval model $M$ is defined as follows.
\begin{equation}
    FD(D,M)=\frac{x}{n}*100
    \label{eq:fidelity}
\end{equation}

 \textbf{Avg. New Words:} Here we compute the average number of new words added by the counterfactual approach for a set of query document pairs. 

\textbf{Avg. Query Overlap:} 
Here we report on an average how many of the words suggested by the counterfactual algorithm come from the query words.

\paragraph{Parameters and Implementation Details}

The details of implementation about retrieval models are shown in Appendix \ref{sec:model_performance}. We employed two popular classical machine learning methods, Logistic Regression (LR) and Random Forest (RF) for the classifier described in Section \ref{sec:classifier}. For Logistic Regression, the learning rate was set to $0.001$. For Random Forest, the number of estimators was set to $100$. 
%
As described in Section \ref{sec:classifier}, all the words present in a document are not used as input to the classifier. We use the top $10$ ($n'=10$) most important words from a document. As described in Section \ref{sec:fv}, we explored three different ways to implement $Imp(d)$ function a) TF-IDF weight based word extraction, b) BERT based keyword extraction \cite{grootendorst2020keybert} and c) Similarity between the BERT representation of query and the document tokens. We found that BERT representation-based similarity computation worked the best for our approach. More details on the implementation of $Imp(d)$ function are shown in Appendix \ref{sec:Imp}.  The value of $K'$ for $Top-K'$ baseline was set to $5$. 
More details on the parameter configuration are shown in Appendix \ref{parameters}.

\section{Results}\label{sec:results}
Table \ref{tab:Results} shows the performance of the counterfactual approach across different retrieval models (i.e. BM25, DRMM, DSSM, ColBERT, MonoT5 and Splade).  We conducted experiments on MS MARCO passage and document ranking dataset and TREC Robust dataset to observe the effectiveness of our proposed explanation approach for different types of documents. Mainly four different observations can be made from Table \ref{tab:Results}. \textbf{\textit{Firstly}},  It can be clearly observed that the CFIR model for each retrieval model has performed better compared to its corresponding query word or top-K' words baseline in terms of Fidelity score(FD). The above-mentioned observation is consistent for both passages and long documents (i.e. in MSMARCO passage, Document and TREC Robust). 
\textbf{\textit{Secondly}}, it can be observed from Table \ref{tab:Results} that mostly CFIR approach provided the highest number of new terms (terms not already present in the documents) as part of the explanation to improve ranking. Consequently, we can say the overall set of explanation terms are more diverse for CFIR approach compared to others. It can also be also observed from Table \ref{tab:Results} that the Fidelity scores are generally better in the MS MARCO passages compared to MSMARCO document and TREC Robust dataset. One likely explanation for this phenomenon is that documents in MSMARCO document and TREC Robust are longer in length compared to passages. Consequently, it is easier for shorter documents to change the ranking compared to longer documents. 
\textbf{\textit{Thirdly}}, another interesting observation from Table \ref{tab:Results} is that the maximum query word overlap by our proposed approach is $68\%$. This implies that the counterfactual algorithm is suggesting new words that are not even present in a query. 
\textbf{\textit{Fourthly}}, the performance of representation learning based retrieval models (i.e. ColBERT, MonoT5) are significantly better than the other models for Fidelity metric. One potential reason can be that, the counterfactual generator suggests words which are similar to the content of the document.  Because of using better embedding representation (BERT \cite{devlin2018bert} and T5 compared to Word2Vec \cite{mikolov2013distributed} in DRMM) these retrieval models give more priority to similar words than other retrieval models.
\begin{figure}
\includegraphics[width=0.40\textwidth]{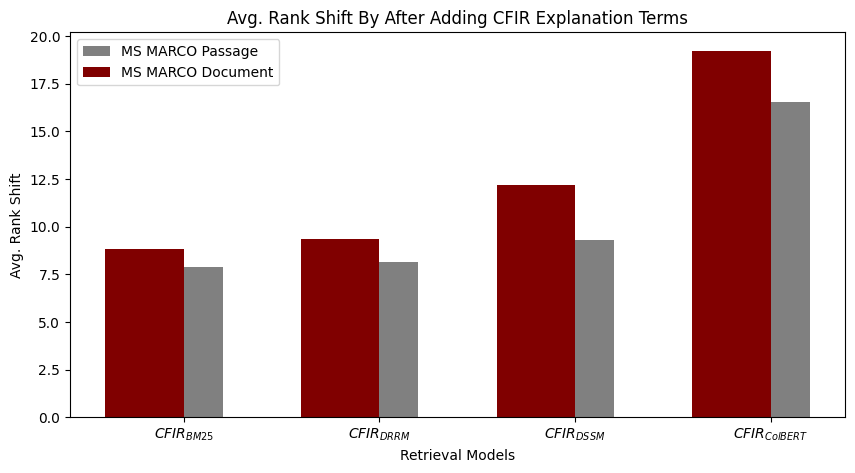}
    \caption{{Average Rank shift by CFIR for BM25, DRMM, DSSM, ColBERT, MonoT5 and Splade}}
    \label{fig:plot_shift}
\end{figure}

Prior work in information retrieval has explored adversarial attacks, where document content or embeddings are perturbed to manipulate rankings with malicious intent \cite{blackbox1, wu:2022:arxiv:prada}. In contrast, the goal of counterfactual explanations is to provide interpretability for IR models by revealing how document rankings can be improved. A key distinction lies in the nature of intervention: adversarial methods typically aim to introduce minimal perturbations often by substituting content, including important terms to preserve the original semantics while deceiving the model. In our case, CFIR explicitly seeks to identify new terms that, when added to a document, improve its rank, thereby highlighting what informative aspects were absent. Replacing important terms is not useful in counterfactual setup, as it fails to address what the document was lacking from the model's perspective. This formulation is particularly relevant for understanding model behavior, including uncovering potentially problematic model preferences (e.g., prior studies have observed gender bias in ranking systems). By identifying helpful additions, such as gendered terms, CFIR can reveal latent model sensitivities. Importantly, unlike adversarial attacks, the size of the added term set is also not constrained in CFIR (Avg. New Words column in Table \ref{tab:Results} shows maximum $16.81$ new words per explanation), as the focus is on explanatory sufficiency rather than minimality.
However, for comparison, we have evaluated the performance of CFIR against the PRADA \cite{wu:2022:arxiv:prada} model which replaces certain words in a document to improve its ranking. Table \ref{pradavscfir} in Appendix \ref{ap:prada} shows that CFIR performs better than PRADA for both ColBERT and MonoT5 in terms of Fidelity score. Table \ref{tab:IllusExIR} in Appendix \ref{ap:examples} shows a sample of example terms extracted by our proposed approach. 
\begin{figure}
\includegraphics[width=0.40\textwidth]{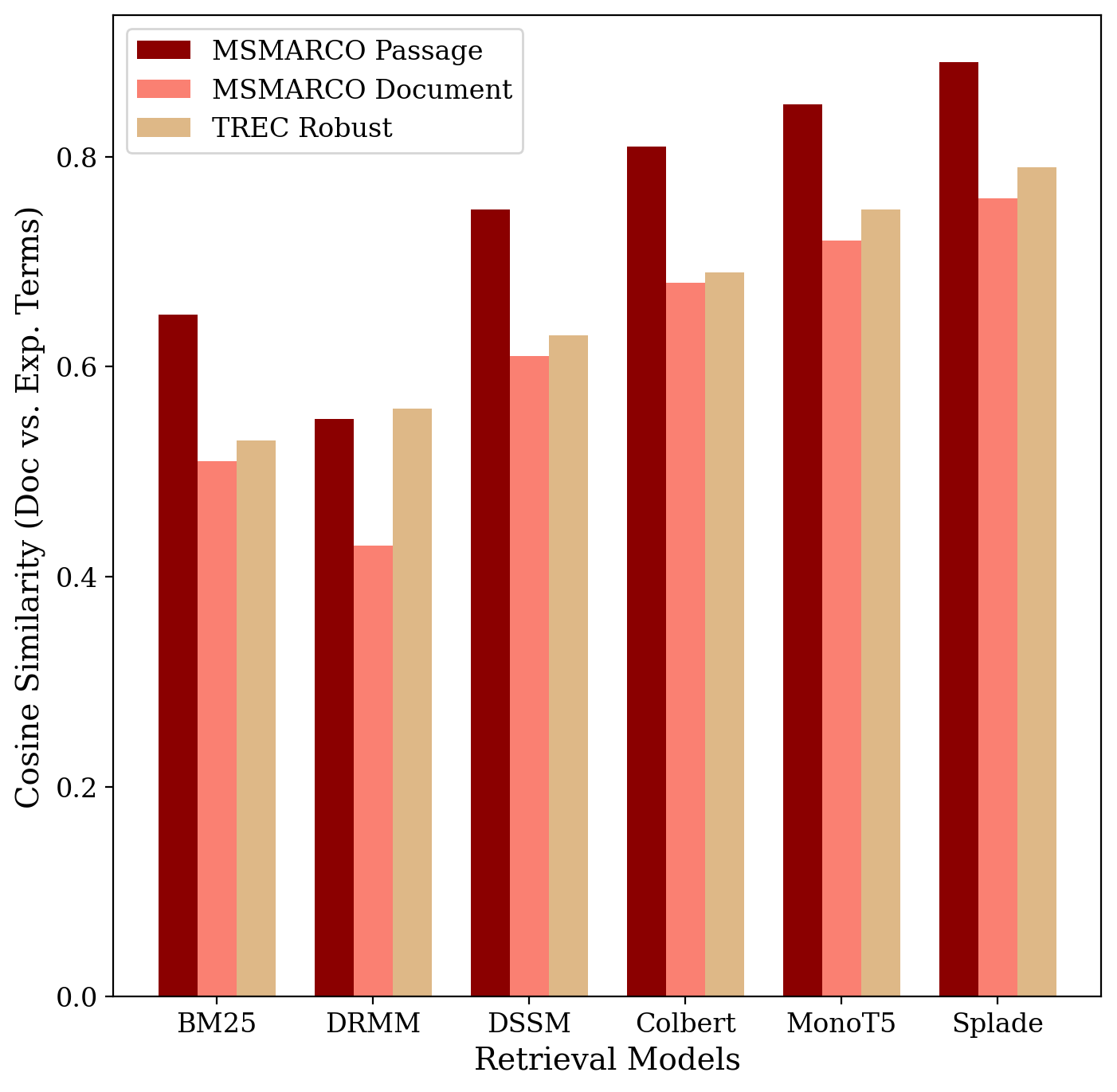}
    \caption{{Average Semantic Similarity between original documents and the corresponding counterfactual explanation Terms for BM25, DRMM, DSSM, ColBERT, MonoT5 and Splade}}
    \label{fig:sim}
\end{figure}
\begin{figure*}[htb]
  \begin{subfigure}[]{}
\includegraphics[width=0.43\textwidth]{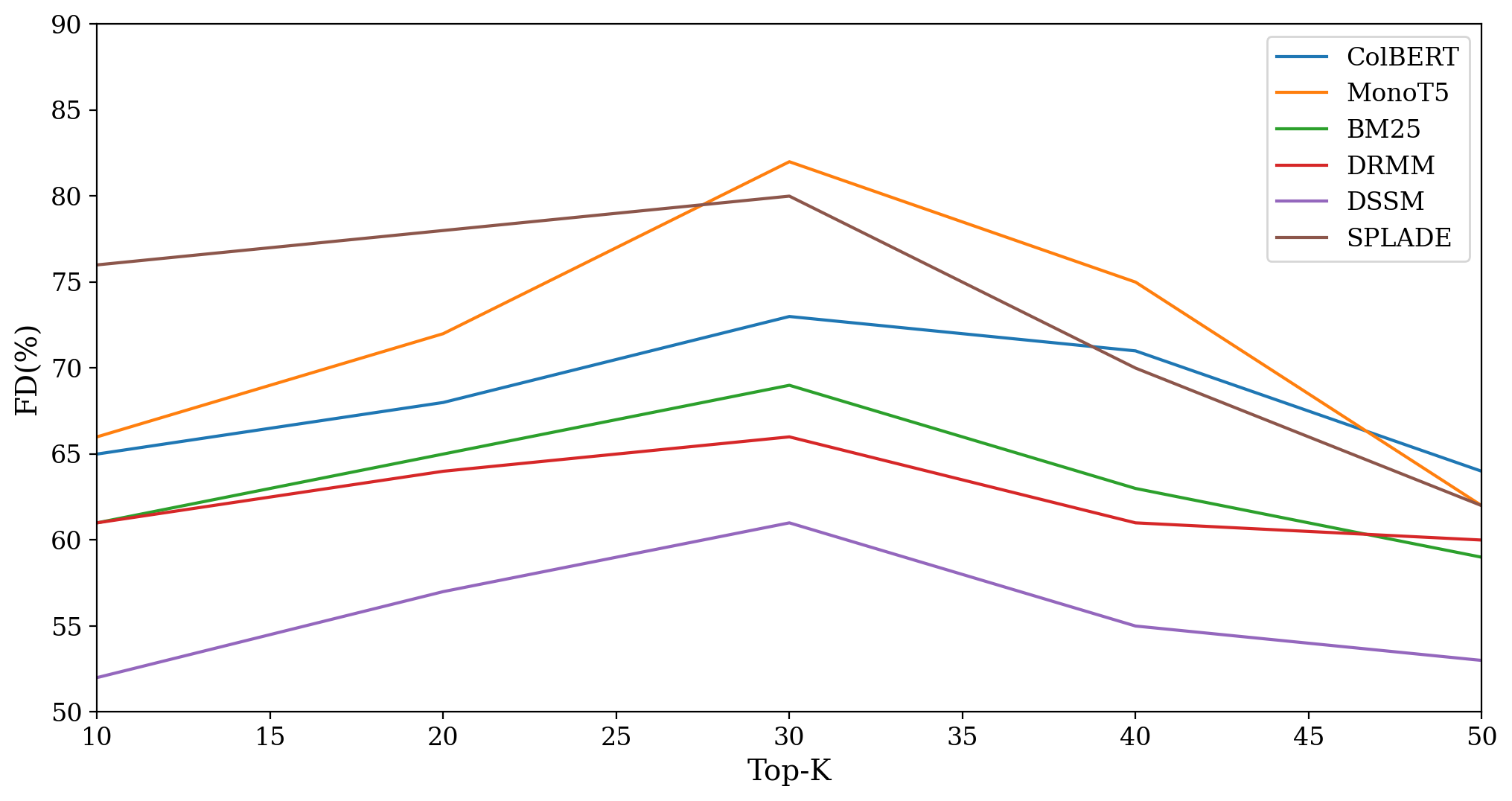}
    \label{fig:my_label}
\end{subfigure}
\begin{subfigure}[]{}
\includegraphics[width=0.43\textwidth]{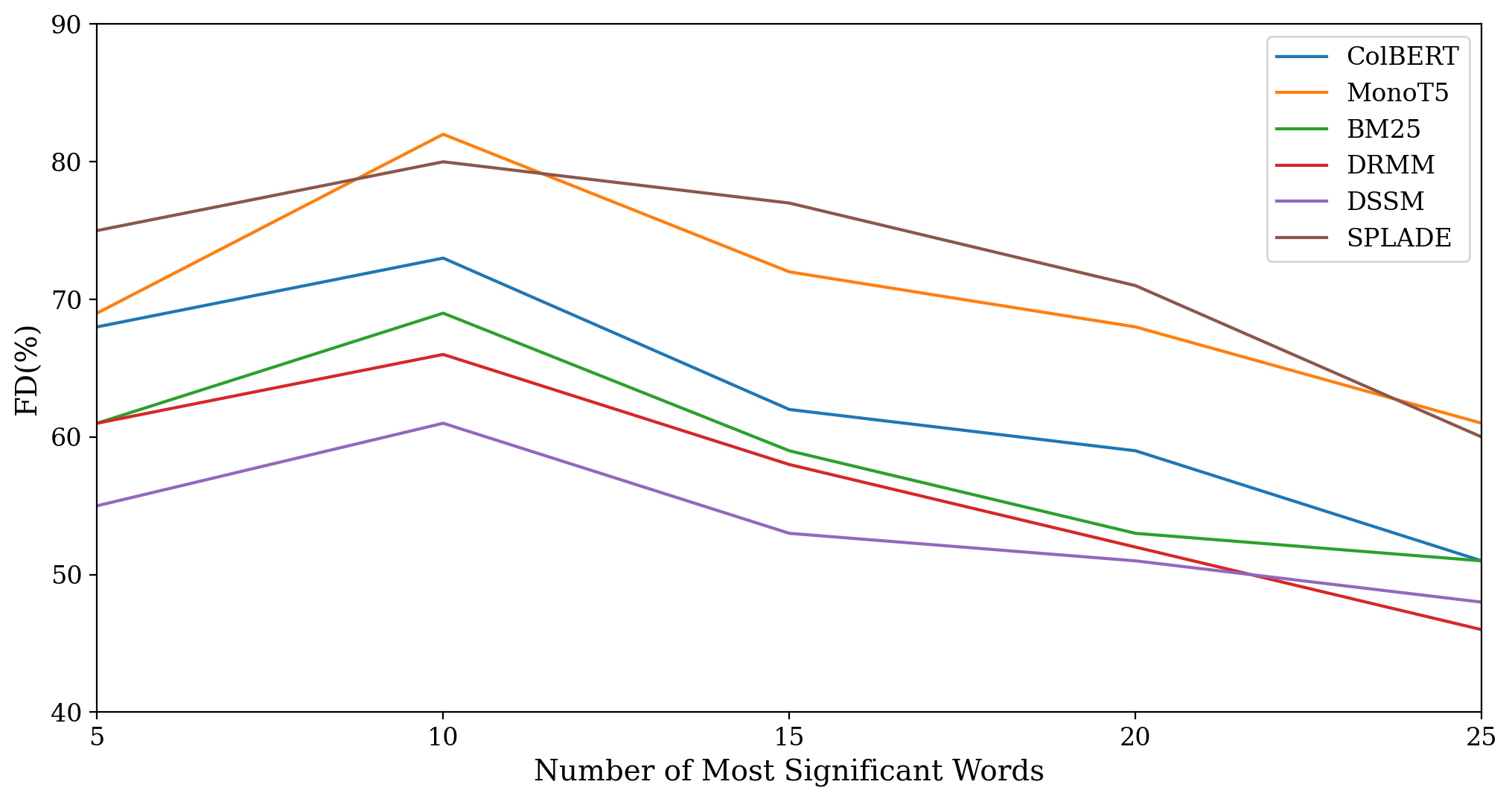}
\label{fig:fig1}
\end{subfigure}
\caption{Counterfactual  Classifier Performance Variance with Top-K and Counterfactual Performance Variance with variation of number of Counterfactuals}
\label{fig:dualfig}
\end{figure*}

\textbf{Further Analysis} Figure \ref{fig:plot_shift} shows the average change in rank after introducing the explanation terms suggested by the CFIR setup. Figure \ref{fig:plot_shift} essentially demonstrates the actionability introduced by the counterfactual explanation terms. The two things to observe from Figure \ref{fig:plot_shift} are firstly, the average rank shift is greater for documents than for passages. Table \ref{tab:Results} shows that ColBERT achieved a significantly higher fidelity score ($16^{th}$ row) and a larger average rank shift compared to the other models, as also seen in Figure \ref{fig:plot_shift}. Figure \ref{fig:sim} shows the average cosine similarity computed between documents and the corresponding explanation terms. For both documents and the explanation terms we use pretrained BERT representations to compute the similarity. It can be observed from Figure \ref{fig:sim} that the cosine similarity for the representation learning based retrieval models (i.e. ColBERT, MonoT5) are higher than the other retrieval models in general.

\textbf{Parameter Sensitivity Analysis}
In Table \ref{tab:Results}, we observed that for most of the retrieval models the performance of the counterfactual explainer follows similar trend both in MSMARCO passage and document dataset (i.e. the best performing model in terms of fidelity score is same in most of the cases). As a result, we conducted parameter sensitivity experiments only on MSMARCO passage dataset. Figure \ref{fig:dualfig} (a) shows the variance in Fidelity score with respect to the K value in Top-K. 
In Figure \ref{fig:dualfig} (b) we show the variance of FD score with respect to the number of most significant words (i.e. $n$) used to construct the document vector. It is clearly visible from Figure \ref{fig:dualfig} (b) that with an increase in the number of counterfactuals, there is a decrease in the performance of the counterfactual classifier. It can be observed that for $n=10$ the best performance is achieved. Intuitively, as the number of words increases, the feature vector grows exponentially, making it challenging to train the classifier effectively. 

\paragraph{Qualitative Evaluation of Explanations}
We conducted a user study involving three researchers with doctoral degrees in IR to estimate the quality of explanations. Each annotator was provided with 30 documents from the MS MARCO passage collection, along with the corresponding queries, ranked lists, and explanation terms generated by CFIR applied to the best-performing model, MonoT5 (shown in Table~\ref{tab:Results}). Further details about the experiment setup is given in Appendix \ref{user_study}. Users were asked to assess the quality of explanations across six dimensions: (a) \textit{Intuitiveness} how intuitive the explanation terms appeared given the query, document, and ranking context, with knowledge of the retrieval model; (b) \textit{Non-intuitiveness} the extent to which explanations felt unexpected or misaligned with the query-document pair; (c) \textit{Query Relatedness} whether the explanation terms were semantically related to the query; (d) \textit{Document Relatedness} whether the explanation terms aligned with the overall topic of the document; (e) \textit{Informativeness} whether the terms were meaningful and content-rich rather than generic or uninformative (e.g. mostly stop words); and (f) \textit{Diversity} whether the explanation terms covered varied semantic aspects. For each aspect the users were asked to put a score between $0$ to $5$. 
\begin{figure}
\includegraphics[width=0.40\textwidth]{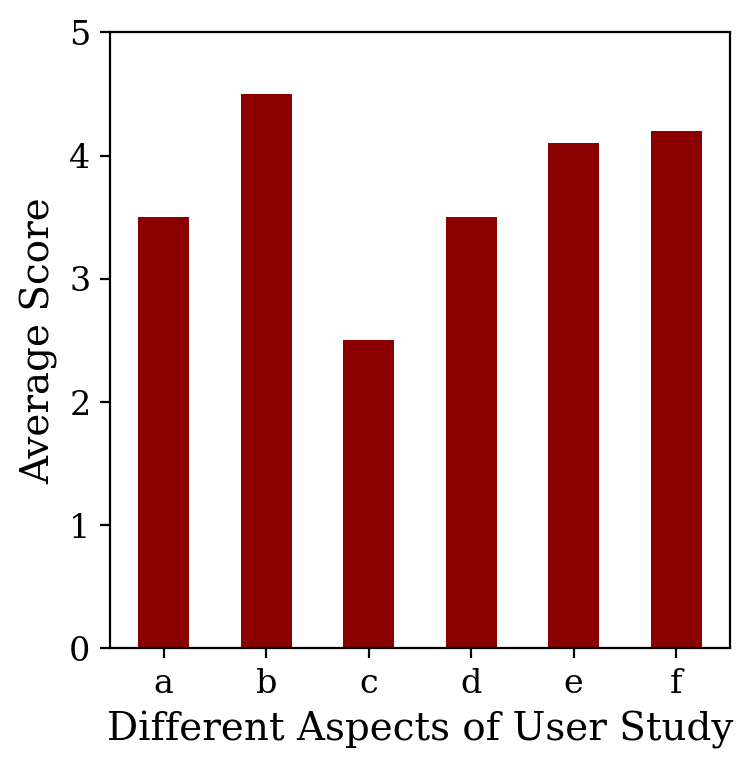}
    \caption{{Qualitative Assessment of Generated Explanations over a) Intuitiveness b) Non-Intuitiveness c) Query Relatedness d) Document Relatedness e)Informativeness f) Diversity)}}
    \label{fig:sim1}
\end{figure}
Figure \ref{fig:sim1} shows that in general the explanation terms are intuitive and more similar to the document topic compared query topic (as expected due to use of document similarity criteria in the loss function in Equation \ref{eq:cfir}). The explanation terms are also quite diverse.  The non-intuitiveness score is quite low which shows that most of explanation terms follow an IR practitioner's intuition. 
\section{Conclusion} \label{sec:conc}
In this paper, we propose a counterfactual setup for a query-document pair and a retrieval model. 
Our experiments show that the proposed approach on an average $70\%$ cases for both in short and long documents could successfully improve the ranking. 
In the future, we would like to explore different explanation units 
for the counterfactual setup.

\section{Limitations}One of the limitations of this work is that we assume that top 10 or 20 words (based on tf-idf weights) within a document play the most
important part in improving the rank of a document. However, theoretically speaking we 
should consider all the words present in a document to determine the most influential words 
for a retrieval model. We have used top tf-idf 
words (Similar to statistical retrieval models) to reduce the computational complexity 
of our experiments and we have seen that increasing the number of top words doesn't affect the performance of the model that much.
\section{Ethical Considerations}
In this work, we have used publicly available
search query log and document collection to
 demonstrate counterfactual explanation. No sensitive data was used in this experiment. As a result of this there is no particular
 ethical concern associated with this work. If
there is any kind of bias present in the search
 log data that effect can be observed within
 our approach. However mitigating that bias
 was beyond the scope of this work
\bibliography{ref}
\appendix
\section{Retrieval Models} \label{ap:rm}
The five different retrieval models used in our experiment are described as follows.

    \textbf{BM25:} BM25\footnote{\url{https://en.wikipedia.org/wiki/Okapi_BM25}} is a statistical retrieval model where the similarity between a query and a document is computed based on the term frequency of the query words present in the document, document frequency of the query words and also the document length. 
   
    \textbf{DRMM:} Deep Relevance Matching Model (DRMM) \citet{guo} is a neural retrieval model where the semantic similarity between each pair of tokens corresponding to a query and a document is computed to estimate the final relevance score of a document.
    
     \textbf{DSSM:} Deep Semantic Similarity Model (DSSM) \citet{huang} is another neural retrieval model which uses word hashing techniques to compute the semantic similarity between a query and a document.

    \textbf{ColBERT:} Contextualized Late Interaction over BERT (ColBERT)  \cite{colbert}, is an advanced neural retrieval model which exploits late interaction techniques based on BERT \cite{devlin2018bert} based representations of both query and document for retrieval.
      
      \textbf{MonoT5:} MonoT5 \cite{nogueira-etal-2020-document} is a sequence-to-sequence model fine-tuned to predict the relevance of a query-document pair.

      \textbf{Splade:}  Splade \cite{splade}(Sparse Lexical and Expansion Model for Information Retrieval) combines the sparse interpretability of traditional IR models (like BM25) with the semantic power of deep learning. Unlike dense retrieval models that rely on vector similarity in embedding space, SPLADE encodes queries and documents into sparse high-dimensional vectors—essentially performing learned term expansions in a way that mimics the inverted index structure used in classic IR systems.

\section{Retrieval Performance of IR Models}\label{sec:model_performance}
We use \citet{pyserini} toolkit for implementing BM25 and MonoT5 and Splade. For DRMM and DSSM, we use the implementation released by the study in \citet{drmm}. For passage ranking  we varied the parameters in a grid search and we took the configuration producing best MRR@10 value on TREC DL \cite{craswell:2021:trec-dl} test set. For both DRMM and DSSM experiments on MSMARCO data, the parameters were set as suggested in \cite{drmmmsmarcodoc}. The MRR@10 values are reported in Table \ref{tab:compExIR} in Appendix \ref{sec:model_performance}. For DRMM and DSSM, we use randomly chosen $100K$ query pairs from the MSMARCO training dataset to train the model. 

The machine used to run counterfactual experiments on retrieval model has 1 A100 GPU and 40 GB memomry. 
\begin{table}[htb]
\footnotesize
\resizebox{0.5\textwidth}{!}
{
\begin{tabular}{|l|c|c|}
\hline
 \rowcolor{gray!30}
&\multicolumn{2}{|c|}{ \textbf{MRR@10}}\\
   \hline
 \rowcolor{gray!30}
 \textbf{Model}& 
  \textbf{MSMARCO Passage}&  \textbf{MSMARCO Document}\\
  \hline
 BM25& $0.1874$ & $0.2184$ \\
DRMM& $0.1623$ & $0.1168$ \\
DSSM& $0.1320$ & $0.1168$\\
ColBERT& $0.3481$ & $0.3469$ \\
MonoT5 &$0.3904$ & $0.3827$\\
Splade &$0.3813$ & $0.3721$\\
\hline
\end{tabular}}
\caption{Retrieval Model Performance on MSMARCO passage and document}
\label{tab:compExIR}
\end{table}
\section{Dataset Statistics} \label{ap:stat}
The dataset statistics for all the experiments are given in  Table \ref{tab:dataset}
\begin{table}[htb]
\footnotesize
\resizebox{\columnwidth}{!}{%
  \begin{tabular}{|l|c|c|c|c|}
   \hline
    \rowcolor{gray!30}
    &&\makecell[cc]{\textbf{MS MARCO} \\ \textbf{Passage}}& \makecell[cc]{\textbf{MS MARCO} \\ \textbf{Document}} & \makecell[cc]{\textbf{TREC} \\ \textbf{Robust}}\\
    \hline
  Query&  Avg Length & 5.9 &  6.9&7.18\\ 
  Document& Avg Length & 64.9 & 1134.2&150.12\\ 
    Query&   \#Instances &100 &  100&100\\ 
 Document&  \#Instances & 500 &  500&500\\ 
    \hline
\end{tabular}%
}
\caption{Dataset Details for Counterfactual Setup}
\label{tab:dataset}
\end{table}

\section{Example of Input and Output to Classifier} \label{ap:classifier}
Given an input query, we employ a LuceneSearcher with MSMARCO Index to retrieve the Top-K documents. The feature vector construction process follows these steps:

For each document, we:
\begin{enumerate}
    \item Extract the top n words based on their Imp(d) values
    \item Construct a vocabulary $V$ as the union of all top 10 words across documents
    \item Note that $|V|$ typically falls in the range of 150-180 words
\end{enumerate}

The feature vector for each document has dimension $|V|$, where each component represents the value from the Imp(d) of the corresponding word from the vocabulary. Formally:

$\text{$d_{vec}$} \in R^{|V|}$


Labels are assigned according to the following criterion:
\[
\text{label} = 
\begin{cases}
1 & \small\text{for top } K \text{ documents} \\ 
0 & \small\text{for remaining documents}
\end{cases}
\]

Example feature vectors and their corresponding counterfactuals generated using \cite{mothilal2020dice} are shown in Table~\ref{fig:exampleclass}. Since |V| is 150 in our experiments, hence in Table \ref{fig:exampleclass} we have only shown the term frequencies of the words present in each document. For other words the terms freaquency values will be zero in $d_{vec}$.
\begin{table}[htb]
\footnotesize
\resizebox{\columnwidth}{!}
{
\begin{tabular}{|l|c|}
   \hline
 \rowcolor{gray!30}
 \textbf{Existing Explanation Methods}& 
 \textbf{Word Overlap}\\
  \hline
 PointWise Explanation \cite{exs}& 21.46\% \\
  ListWise Explanation \cite{IXS}& 9.57\% \\
    \hline
\end{tabular}}
\caption{Comparison of CFIR with Existing ExIR Approaches}
\label{tab:compExIR}
\end{table}
  \begin{table*}[htb]

\resizebox{\textwidth}{!}
{
\scriptsize
\begin{tabular}{|l|c|}

   \hline
     \rowcolor{gray!30}
 \textbf{ docID}&  \textbf{Feature Vector}  \\
  \hline
3686955 & [prohibition:2.0, amendment:2.0, under:1.0, dwindled:1.0, eighteenth:1.0, repeal:1.0, repealed:3.0, states:1.0, 1933: 1.0, ratification: 1.0]\\
6159679  & [membrane:5.0, lipids:3.0,  remainder:2.0, proteins:3.0, biochemical:2.0, 80:2.0, role:2.0, percent:2.0]\\
5217641 & [waves:6.0, transverse:5.0, electromagnetic:3.0, oscillations:2.0, vibrations:2.0, travel:2.0, radiation:2.0, angles:2.0, transfer:2.0, types:3.0]\\
    \hline
\end{tabular}}
\caption{Sample Feature Vector Corresponding to three different documents}
\label{fig:exampleclass}
\end{table*}
 \section{Scalability Issues}
There can be concerns regarding the feasibility of training a classifier per document. To address this, we propose and evaluate an alternative and more efficient strategy in which a single classifier is trained per query, rather than per document. Concretely, for a given query, we train one classifier using: (i) all documents for which explanations are required (let this number be $x$); (ii) their nearest neighbors, which contribute to the non-relevant document set; and (iii) the top-k retrieved documents. To balance the number of relevant and non-relevant training instances, we construct the non-relevant set with a total size of 2k, where we sample 2k/x nearest neighbors from each document for which explanations are generated.

The results of this per-query training strategy are reported in Table \ref{tab:scalability}. As shown, this substantially faster approach achieves performance comparable to that reported in Table 2, where a separate classifier was trained for each document. This demonstrates that our method remains effective while significantly reducing the computational overhead, directly addressing the reviewers’ feasibility concerns.
 \begin{table*}[htb]
   \footnotesize
\resizebox{\linewidth}{!}
{
\begin{tabular}{|lc|c|c|c|c|c|c|c|c|c|}
    \hline
     \rowcolor{yellow!25}
   \multicolumn{2}{|c|}{\textbf{Model Description}}& \multicolumn{3}{c|}{\textbf{MS MARCO Passage}} &\multicolumn{3}{|c|}{\textbf{MS MARCO Document }} &\multicolumn{3}{|c|}{\textbf{Trec Robust }} \\
     \hline
     \rowcolor{gray!30}
        \textbf{Retrieval Model} & \textbf{Classifier} &  \textbf{FD(\%)} & \makecell[c]{\textbf{Avg. New} \\ \textbf{Words}} & \makecell[cc]{\textbf{Avg. Query} \\ \textbf{Overlap}}&    \textbf{FD(\%)} & \makecell[cc]{\textbf{Avg. New} \\ \textbf{Words}} & \makecell[cc]{\textbf{Avg. Query} \\ \textbf{Overlap}}&\textbf{FD(\%)} & \makecell[cc]{\textbf{Avg. New} \\ \textbf{Words}} & \makecell[cc]{\textbf{Avg. Query} \\ \textbf{Overlap}}\\
    \hline
$CFIR_{MonoT5}$ &LR  &\underline{\textbf{81.16}}\%
& 12.45 & 63\% & \underline{\textbf{73}}\%&
\textbf{13.13}&63\%& \underline{\textbf{74}}\% & \textbf{10.45}
& 67\%\\
\hline
\hline
$CFIR_{Splade}$ &RF &78\%
& 11.23 & 62\% & 69\%&
12.11&60\%& 71\% & 9.81
& 65\%\\

$CFIR_{Splade}$ &LR  &\textbf{76.92}\%
& 12.15 & 63.4\% & \underline{\textbf{68}}\%&
\textbf{11.33}&64\%& \underline{\textbf{70.11}}\% & \textbf{8.91}
& 67\%\\
\hline
\end{tabular}}
\caption{CFIR model Performance for MonoT5 and Splade in MSMARCO Passage and Document  Collection and TREC Robust. The Best Performing Counterfactual Explanation Method for every retrieval model is boldfaced; the overall best performance across all rows is underlined. All the results reported in Table \ref{tab:Results} are statistically significant with $p < 0.05$.}
\label{tab:scalability}
\end{table*}
\section{Existing EXIR approaches vs. CFIR} \label{sec:exp}
The existing literature aims to explain the significance of a document, a set of documents, or a pair of documents through various explanation methods. Nonetheless, our proposed approach diverges fundamentally from prior work in that we seek to demonstrate how the absence or frequency of certain tokens impacts document relevance. In this section, we examine whether there is any intersection between the two sets of tokens described earlier.
\paragraph{\textbf{Pointwise Explanation Approach}}
As outlined in Section \ref{sec:expir}, existing pointwise explanation methods elucidate why a specific document aligns with a given query within a retrieval model. Similarly, our proposed approach operates on individual documents and queries, albeit with a distinct objective. Here, we analyze the overlap between the explanations generated by the pointwise explanation method and those derived from our model, as presented in Table \ref{tab:compExIR}. This comparison was conducted on 50 pairs of documents.

\paragraph{\textbf{Listwise Explanation Approach}}
In Section 2, it is explained that listwise explanations typically aim to demonstrate the relevance of a list of documents to a given query. In listwise setup, one set of explanation terms are extracted for a list of documents, a query, and a retrieval model. Conversely, in our approach, we generate distinct explanations for each query-word pair. Therefore, to compare listwise explanations with our method, we aggregate all individual explanations obtained for each document-query pair in the list to create a unified explanation set for the entire list corresponding to a query. The resulting overlap is presented in Table \ref{tab:compExIR}.
\begin{table}[htb]
\footnotesize
\resizebox{\columnwidth}{!}
{
\begin{tabular}{|l|c|}
   \hline
 \rowcolor{gray!30}
 \textbf{Existing Explanation Methods}& 
 \textbf{Word Overlap}\\
  \hline
 PointWise Explanation \cite{exs}& 21.46\% \\
  ListWise Explanation \cite{IXS}& 9.57\% \\
    \hline
\end{tabular}}
\caption{Comparison of CFIR with Existing ExIR Approaches}
\label{tab:compExIR}
\end{table}
\begin{table*}[htb]

\resizebox{\textwidth}{!}
{
\scriptsize
\begin{tabular}{|l|c|c|c|}

   \hline
     \rowcolor{gray!30}
 \textbf{ Retrieval Model}&  \textbf{Query Text} &docId&\textbf{Explanation Terms} \\
  \hline
 DRMM & What law repealed prohibition ?&  3686955 &\makecell[cc] {working, strict, Maine, 1929, law, resentment, New York City,\\ Irish, immigrant, prohibition, repeal, fall, Portland, temperance, riot, visit }\\
\hline
 DSSM & What is the role of lipid in the cell?&6159679& phospholipid, fluidity, storage, triglyceride, fatty receptor \\
 \hline
 ColBERT  &what type of wave is electromagnetic? & 5217641&directly ,oscillations, medium, wave, properties, speed\\
 \hline
MonoT5  &what is a caret? & 6338711&display, diamond, weight\\
 \hline
Splade  &which vitamins help heal bruises? & 3465680&minerals, body, eat, cut\\
    \hline
\end{tabular}}
\caption{CFIR explanation terms for DRMM, DSSM, ColBERT, MonoT5 and Splade in MS MARCO passage.}
\label{tab:IllusExIR}
\end{table*}
\section{Counterfactual Optimization Framework} \label{ap:cfsetup_appendix}
The different parts of Equation \ref{eq:cfsetup} are described here. The $yloss$ in Equation \ref{eq:cfsetup} is a hinge loss function as defined in Equation \ref{eq:loss}. In Equation \ref{eq:loss} $z$ is $-1$ when $y = 0$ otherwise, $z=1$. $logit(f_{\{M,q\}}(c_i))$ is the logit values obtained from the classifier ($f_{\{M,q\}}$)  when the counterfactual $c_i$ is given as input. 

\begin{equation}
    yloss= max(0, 1-z*logit(f_{\{M,q\}}(c_i)))
    \label{eq:loss}
\end{equation}

The distance function ($dist(c_i,d)$) in Equation \ref{eq:cfsetup} is computed using the formula given in Equation \ref{eq:madp}.  In Equation \ref{eq:madp}, $V$ represents the vocabulary set used to represent the document vectors ($d_{vec}$). In Equation \ref{eq:madp}, the value of $I$ is equal to $1$ if the corresponding term is present in both the counterfactual input $c$ and the original input $d$, otherwise it is set to $0$.
\begin{equation}
dist(c,d)=\sum_{p=1}^{V}I (c_p \neq d_p)
\label{eq:madp}
\end{equation}
The diversity in above equation is defined by the formula described in Equation \ref{eq:diversity}. In equation \ref{eq:diversity}, $K_{i,j}$ is equal to $\frac{1}{1+dist(c_i,c_j)}$. $dist(c_i,c_j)$  calculates the distance between two counterfactuals $c_i$ and $c_j$.
\begin{equation}
    div(c_1,\ldots,c_k)=\sum_{i,j}det(K_{i,j})
    \label{eq:diversity}
\end{equation}

\section{Parameters for Counterfactual Setup}\label{parameters}
The value of $\lambda_1$ and $\lambda_2$ is set to $1$ and $0.5$ respectively in Equation \ref{eq:cfsetup}. The value of $k$ in Equation \ref{eq:cfsetup} is set to $k=3$. In all our experiments in Table \ref{tab:Results}, we have observed that for $K=3$ and onward we have always found a counterfactual explanation for each query-document pair where only words were added for the desired counterfactual outcome.

\section{Comparison with Credence\cite{Rorseth_2023}}
The counterfactual method proposed in CREDENCE\cite{Rorseth_2023} identifies sentences whose removal leads to a decrease in the document’s rank. In contrast, our work focuses on the complementary setting — identifying interventions that can increase a document’s rank. Second, the method in CREDENCE relies on the intuition that removing any sentence containing query terms lowers the ranking score. This is closely related to our query-word baseline, where we instead add missing query terms to the document to improve its rank. Third, Credence employs a heuristic procedure in which each candidate explanation document is repeatedly reranked and then verified to determine whether it is a valid explanation. In contrast, we introduce an explicit objective function that directly guides the optimization, thereby avoiding the expensive reranking loop required by heuristic approaches. However, based on your suggestion we have now aslo done direct comparison of CREDENCE with our proposed approach. We have used the reverse heuristics of CREDENCE to make this method comparable with our approach. We have added sentences containing query terms to the document to improve its rank. The results are as follows.
\begin{table*}[htb]
\footnotesize
\centering
\resizebox{0.8\textwidth}{!}
{
\begin{tabular}{|l|c|c|c|c|c|c|c|c|c|}
    \hline
     \rowcolor{yellow!25}
  & \multicolumn{3}{c|}{\textbf{MS MARCO Passage}} &\multicolumn{3}{|c|}{\textbf{MS MARCO Document }} &\multicolumn{3}{|c|}{\textbf{Trec Robust }} \\
     \hline
     \rowcolor{gray!30}
        \textbf{Model}  &  \textbf{FD(\%)} & \makecell[c]{\textbf{Avg. New} \\ \textbf{Words}} & \makecell[cc]{\textbf{Avg. Query} \\ \textbf{Overlap}}&    \textbf{FD(\%)} & \makecell[cc]{\textbf{Avg. New} \\ \textbf{Words}} & \makecell[cc]{\textbf{Avg. Query} \\ \textbf{Overlap}}&\textbf{FD(\%)} & \makecell[cc]{\textbf{Avg. New} \\ \textbf{Words}} & \makecell[cc]{\textbf{Avg. Query} \\ \textbf{Overlap}}\\
    \hline
$CREDENCE_{BM25}$	& 64\% & 	8.32& 	63\%& 	52\%& 	11.51& 	55\%& 	56\%& 	12.21& 	54\%\\
$CREDENCE_{DRMM}$& 	68\%& 	9.11& 	59\%	& 58\%	& 10.32	& 42\%	& 61\%& 	13.52	& 42\% \\
$CREDENCE_{DSSM}$& 	60\%	& 14.12	& 57\%& 	51\%	& 16.51& 	60\%& 	54\%	& 8.51	& 61\% \\
$CREDENCE_{ColBERT}$	& 71\%	& 11.21& 	54\%& 	69\%& 	10.52& 	52\%& 	68\%& 	12.56& 	53\% \\
$CREDENCE_{MonoT5}$& 	71\%	& 11.23&	62\%	& 70\%	& 14.12	&53\%& 	72\%	& 10.91& 	64\% \\
$CREDENCE_{Splade}$	& 75\%	& 9.72& 	65\%& 	70\%& 	10.35	& 62\%	& 68\%	& 9.21	& 59\% \\

    \hline
\end{tabular}}
\caption{Comparison of CFIR with Existing Counterfactual Approach \cite{Rorseth_2023}}
\label{tab:compExIR}
\end{table*}
 
 \section{Example of Counterfactuals Produced by CFIR Setup} \label{ap:examples}
 The words shown in Table \ref{tab:IllusExIR} have improved the ranking of a docID with respect to the queries shown.

\section{Classifier Accuracy}
The accuracy of the counterfactual classifier build in our setup is described in Table \ref{cc_classifier}.

\begin{table}[]
\centering
\resizebox{0.5\columnwidth}{!}
{\begin{tabular}{|l|c|}
\hline
\rowcolor{gray!30}
 \textbf{Method}& Accuracy  \\
  \hline
$Classifier_{BM25}$ &  81\%\\
$Classifier_{DRMM}$	 & 85\%\\
$Classifier_{DSSM}$	 & 82\%\\
$Classifier_{ColBERT}$ &	84\% \\
$Classifier_{MonoT5}$	 & 86\% \\
$Classifier_{Splade}$	 & 88\% \\
  \hline
 \end{tabular}}
 \caption{Accuracy of the classifier used for counterfactual explanation.}
 \label{cc_classifier}
 \end{table}
\section{Adversarial Attacks vs. Counterfactual Explanation} \label{ap:prada}
Here we show the performance of our proposed counterfactual explanation approach with an existing adversarial model named PRADA \cite{wu:2022:arxiv:prada}. We use the MSMARCO passage dataset as the target corpus. We use same test set (as described in Table \ref{tab:dataset}) as used in the first column of Table \ref{tab:Results} in this experiment. Table \ref{pradavscfir} shows the results in terms of Fidelity score.
\begin{table}[]
\centering
\resizebox{0.9\columnwidth}{!}
{\begin{tabular}{|l|c|c|}
\hline
\rowcolor{gray!30}
 \textbf{ Retrieval Model}&  \textbf{FD in PRADA} &\textbf{FD in CFIR} \\
  \hline
  ColBERT &  74\%  & 75\%\\
  MonoT5  &  80\% & 82\%\\
  \hline
 \end{tabular}}
 \caption{Performance of CFIR vs. Adversarial Attack Model PRADA \cite{wu:2022:arxiv:prada}}
 \label{pradavscfir}
 \end{table}
 \section{Implementation of Imp(d)}\label{sec:Imp}
We explored three ways to compute the top $n$ words from each document. Each one of them is described as follows.

\paragraph{TF-IDF Approach:} In this approach we choose top $n$ words from a document based on their TF-IDF weight.

\paragraph{KEYBERT Approach:} In this approach we use the model proposed in \cite{grootendorst2020keybert} to extract keywords from a string.

\paragraph{BERT-Based Similarity(BERTSim):} In this approach we compute the similarity between the BERT based representation of the query text and each token of the document and then we sort all the tokens based on the similarity.

Table \ref{impd} shows the performance of the above-mentioned three approaches in MSMARCO passage dataset and ColBERT retrieval model. $n=10$ for the experiments shown in Table \ref{impd}. From Table \ref{impd}, we can conclude that the BERT-based similarity approach works the best for the $Imp(d)$ function. hence for all the results reported in Table \ref{tab:Results}, we use the BERTSim approach in the $Imp(d)$ function. 
\begin{table}[]
\centering
\resizebox{0.5\columnwidth}{!}
{\begin{tabular}{|l|c|}
\hline
\rowcolor{gray!30}
 \textbf{ $Imp(d)$ Approach}& FD  \\
  \hline
 TFIDF &  74\%\\
  KeyBERT  & 70\%\\
BERTSim  &   \textbf{75}\%\\
  \hline
 \end{tabular}}
 \caption{Performance of Different Approaches in $Imp(d).$}
 \label{impd}
 \end{table}

 \section{User Study}\label{user_study}
 In the user study we didn't record any personal information of any of the users. We only recorded their judgment about the output of the proposed methodology for the study. We have also used data which is publicly available for IR research. Hence no ethics approval was required for the study. All the researchers were made aware of the of the use of their assessment in this research.

\end{document}